\newcommand{\be}[0]{\begin{equation}}
\newcommand{\ee}[0]{\end{equation}}
\newcommand{\bea}[0]{\begin{eqnarray}}
\newcommand{\eea}[0]{\end{eqnarray}}
\documentstyle[12pt,epsfig]{article}
\begin{document}
\large
\hfill\vbox{\hbox{DTP-97/02}
            \hbox{March, 1997}}
\nopagebreak

\vspace{2.0cm}
\LARGE
\centerline{\bf Unquenching the scalar glueball} 
\vspace{0.8cm}
\begin{center}
\large

M. Boglione $^{1,2}$

\vspace{0.5cm}
and
\vspace{0.5cm}

\large

M.R. Pennington $^1$

\vspace{0.6cm}
\normalsize
\begin{em}
$^1$ Centre for Particle Theory, University of Durham\\
     Durham DH1 3LE, U.K.\\

\vspace{0.4cm}

$^2$ Dipartimento di Fisica Teorica, Universit\`a di Torino and \\
      INFN, Sezione di Torino, Via P. Giuria 1, 10125 Torino, Italy\\
\end{em}

\vspace{1.3cm}

\end{center}
\normalsize

\vspace{0.45cm}

{\leftskip = 1.9cm 
 \rightskip = 1.9cm
\centerline {Abstract}

\noindent
Computations in the quenched approximation on the lattice predict the lightest  
glueball to be a scalar in the $1.5-1.8$ GeV region. Here we calculate the
dynamical effect the coupling to two pseudoscalars has on the mass, width and
decay pattern of such a scalar glueball. These hadronic interactions allow
mixing with the $q \overline q$ scalar nonet, which is largely fixed by the 
well-established $ K_0^*(1430)$. This non-perturbative mixing means that,
if the pure gluestate has a width to two pseudoscalar channels of $\sim100$
MeV as predicted on the lattice, the resulting hadron has a width to these 
channels of only $\sim30$ MeV with a large $\eta\eta$ component.
Experimental results need to be reanalyzed in the light of these predictions  
to decide if either the $f_0(1500)$ or an $f_0(1710)$ coincides with this 
dressed glueball.
\par}
\newpage
\baselineskip=7mm
\parskip=3mm
%
%
\noindent
QCD without quarks, the gauge theory of gluon interactions, predicts a
spectrum of hadrons quite unlike the world experiments so clearly reveals.
The imprint of this pure gauge world would be a spectrum of glueballs, the
lightest of which would be stable with scalar quantum numbers. 
Lattice calculations have now reached sufficient precision to predict this
scalar glueball has a mass of $1740 \pm 71$ MeV \cite{sexton} or 
$1550 \pm 50$ MeV \cite{bali}
depending on how the lattice data are analysed. 
More recently, using improved lattice
actions, Morningstar and Peardon \cite{morn} give results that allow a
central mass of
$1600$ MeV to be deduced. Moreover, the {\it IBM} group \cite{ibm} have 
calculated its coupling to two pseudoscalars
would give a width of $100$ MeV for the scalar, if it could decay to these.
How this state of the quarkless world would appear in the real world is
what this letter is about. 

\noindent
For more than thirty years we have understood that (most) hadrons are
closely connected to the states of underlying quark multiplets: the vector
and tensor mesons and the baryon octet and decuplet being the best known 
examples. The properties of these states are determined by the bound state
dynamics of quarks. This fixes their masses and decays. Thus, it is natural
that the $f_2(1270)$ and $f_2(1525)$ decay predominantly to $\pi \pi$ and 
$K \overline K$ respectively. 
Indeed, it is the composition of the underlying state that
determines its decays, while the masses of the open channels are incidental.
However, as emphasised by Tornqvist \cite{torn}, scalar mesons may be
rather different.
They couple far more strongly and are far more sensitive to the opening of 
thresholds, particularly those with S-wave interactions. 

\noindent
In the quark model we build $0^{++}$ mesons by having a $q \overline q$
system with spin one and a unit of orbital angular momentum. The lightest
of these is expected in the $1-2$~GeV region, in which the scalar
glueball also occurs. The isospin zero $q \overline q$ members are expected
to mix with the gluestate to give the hadrons we observe. 
While a simple quark model of mixing using wholly perturbative methods may
be appropriate for other quantum numbers, the properties of scalars makes
such calculations far too simplistic. Instead the mixing is highly 
non-perturbative and requires a more detailed discussion of the hadron 
propagators. The natural vehicle for this is the appropriate 
Schwinger-Dyson equation. To calculate this, we assume the coupling to two 
pseudoscalar channels controls the dynamics and hence the mixing.

%
%
\vspace{0.3cm} 

\noindent {\bf The calculation.} We begin with the quenched approximation, 
which delivers a bare state of mass
$m_0$ with point couplings $g_i$ to each pseudoscalar channel. Reflecting the
spatial extent of hadrons, these couplings are multiplied by form factors
$F(k_i^2)$, where $k_i$ is the channel's c.m. 3-momentum and 
\be 
F(k_i^2)=\exp [-k_i^2/(2k_0^2)] \; ,
\label{F}
\ee
$k_0$ is related to the interaction radius, which is taken to be between $0.5$
and $1$ fm. As a consequence of chiral dynamics, the couplings to two
pseudoscalars have an Adler zero at $s_{A,i}$, so the full coupling
$G_i$ is given by
\be
G_i^2 = g_i^2 \, (s - s_{A,i}) \, F^2(k_i^2) \; .
\label{G}
\ee
The effect of unquenching is to dress the bare bound state propagators. This
gives them imaginary parts that are the prerequisite for decay. The Dyson
summation of Fig.~1 gives
\be
m^2(s) = m_0^2 + \Pi (s) \; ,
\ee
where $\Pi (s)$ is assumed to be dominated by the two pseudoscalar meson
loops. 

%
\begin{figure}[b]
\mbox{~\epsfig{file=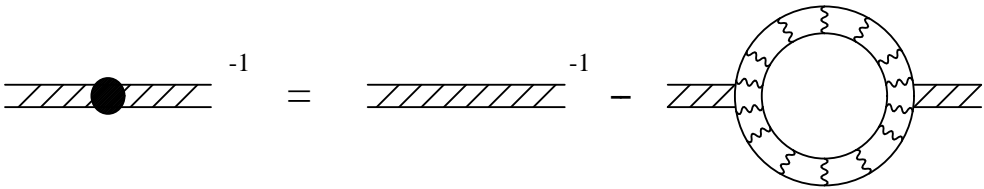,angle=0,width=13cm}}
\caption{The bare bound state propagator is dressed by hadronic
interactions. The dot signifies the dressed propagator. The loop is of 
$q \overline q$ pseudoscalars: the wiggly lines are to enphasize these too
are bound states.}
\end{figure}
%

\noindent
Its imaginary part is simply related to the couplings of Eq.~(\ref{G}) 
by the appropriate phase-space factor
\be
{\rm Im} \Pi _i (s) = - \rho _i (s) \, G_i^2 (s) \ \theta(k_i^2)\; ,
\ee
where $\rho _i (s) = k_i(s)/ \! \sqrt{s}$ .
The real part of $\Pi (s)$ is now fixed by an unsubtracted dispersion relation
-- unsubtracted because hadrons have a spatial extent making only a finite 
number of hadronic channels relevant. Thus
\be
{\rm Re} \Pi (s) = \frac{1}{\pi} \; \; {\cal P} \! \! \int _{s_{th}} 
^{\infty} \! ds' \; \frac{ {\rm Im} \Pi (s')}{s'-s} \; ,
\label{Re}
\ee
with 
\be
{\rm Im} \Pi (s) = \sum _i {\rm Im} \Pi _i (s) \; ,
\label{Im}
\ee
where we sum over all two pseudoscalar modes, e.g. $K \pi ,\, K \eta ,
\, K \eta'$ for the $I=1/2$ channel.
The effect of these meson loops is to produce a propagator
\be
\Delta = \frac{1}{m^2(s) - s} = \frac{1}{m_0^2 + \Pi (s) -s} \; ,
\label{P}
\ee
where the pole is now in the complex energy plane (with reflections on each
unphysical sheet generated by the thresholds to which the state couples).
It is the position of this pole at $s = m_{pole}^2 - i m_{pole} \Gamma_{pole}$
that defines the mass, $m_{pole}$, and width, $\Gamma_{pole}$, of the hadron.

\noindent This is the formalism that applies in the $I=1/2$ and $1$ scalar 
channels,
where the ground state quark model gives one bare `seed'. However, in the 
$I=0$ channel we have more bare states: two ground state $q \overline q$'s
and the
glueball. The philosophy here is that all mixing occurs through communicating
hadronic final states. Thus the bare $q \overline q$ nonet is assumed to be 
ideally mixed.
This is in keeping with the notion that the $\phi$, for example, decays to
$\rho \pi$ through hadron interactions, eg. $\phi \to K \overline K \to \rho
\pi$, and not through any non-hidden-strangeness at the bare level.
With $3$ bare states ($n{\overline n}$, $s{\overline s}$, $gg$) of the same
quantum numbers, the propagator of
Eq.~(\ref{P}) becomes a $3 \times 3$ matrix. 
The hadrons are then the eigenstates of this matrix. 
To determine these requires a diagonalization. 
This diagonalization accounts for the mixing between states
 of the same quantum numbers.

%

\vspace{0.3cm} 

\noindent
{\bf The input} for the bare seeds is as follows.
\begin{itemize}
\item
There is an ideally mixed $q \overline q$ multiplet. The mass of the
non-strange members is $m_0(n \overline n)$, and the mass splitting is 
determined by the
extra mass of the strange quark, where $\Delta m_s \simeq 100 $ MeV from
standard  phenomenology. The couplings of these bare bound states to
two pseudoscalars are assumed to be related by $SU(3) _f$ symmetry, so that in 
fact there is effectively only one overall coupling for the nonet, $\gamma$. 
Similarly, the cut-off $k_0$ in the form factor of Eq. (\ref{F}) is assumed
universal. 
The parameters $m_{0}$, $\gamma$ and $k_{0}$ are to be determined.
\item
The quenched gluestate has mass $m_0(gg)$ specified by lattice calculations.
The coupling $\gamma '$ is arranged to give a width of the unmixed glue
state, to the sum of the two pseudoscalar channels we consider, of 
$(108 \pm 29)$ MeV as computed by the {\it IBM} group~\cite{ibm}. The
coupling to individual channels is assumed to be according to one of two
schemes, either\\
(a) the bare state is an $SU(3) _f$ singlet, or \\
(b) the bare state has couplings as found by the {\it IBM} \\
    \makebox[0.4in][r] group~\cite{ibm} on the lattice.
\end{itemize}
%
\noindent The size of the effect of turning on hadron loops can be understood
qualitatively. For S-wave decays, each threshold produces a jump, Eq.~(4),
in the imaginary part of the mass function directly related to the strength
of the coupling to the opening threshold with a corresponding cusp in the real 
part. If the coupling strength gives a
width of $100$ MeV, for example, then this may shift the mass of any bare
state that is within $100$ MeV or so of this threshold by roughly $100$ MeV
and generally towards this threshold. This naturally has the effect of
making the scalar mesons appear close to strongly coupled 
thresholds~\cite{Corsica}, potentially attracting the $a_0(980)$ and 
$f_0(980)$ to $K \overline K$ threshold, for example, if the parameters are
suitable. 

\vspace{0.8cm}

\noindent {\bf The ${\bf I=1/2}$ and ${\bf 1}$ sectors.} 
Though all the light scalars 
feature in our discussion, the major experimental
ingredient in determining $m_0, \, \gamma, \, k_0$ is the well established 
$K _0 ^*(1430)$, which from the $PDG$ tables~\cite{PDG} has
\be
m = (1429 \pm 6)\, {\rm MeV} ; \; \Gamma =\,(287 \pm 23)\, {\rm MeV} .
\label{K}
\ee
That this $I=1/2$ state is a member of the lightest $q \overline q$ nonet is
uncontroversial. In contrast, whether the $a_0(980)$, or a possible $a_0(1430)$,
belongs to this same multiplet is {\it a priori} not so clear.

\noindent In a closely related analysis \cite{torn}, Tornqvist has
determined the parameters $m_0,
 \, \gamma, \, k_0$ with great precision from the $K \pi$ phase shift that
reflects the presence of the $K_0 ^*(1430)$. To do this he assumes  the $K
\pi$ scattering amplitude is pole dominated. This not only requires the
denominator of any pseudoscalar scattering amplitude to be given by the
propagator of Eq. (\ref{P}), but assumes the numerator has a most specific
form. This is a very special unitarization that is far from the most general,
in particular it lacks crossing-symmetry.
We do not enter here into the controversy~\cite{PRL} about what effect 
Tornqvist's strong
assumptions have on the existence of a light $\sigma$ resonance, suffice it
to say that the treatment by Roos and Tornqvist \cite{roos} neglects the 
cross-channel dynamics that would appear in the numerator of an $N/D$
analysis.  Here it is only the $D$-function that enters our discussion.

\noindent
Since we consider just the denominators of the scalar amplitudes,
 we find a
range of values of the parameters $m_0,\gamma, k_0$ to be consistent with
the strange state's mass and width, Eq.~(\ref{K}) -- a range that includes 
Tornqvist's numbers.
With central values we show in Fig.~2 the real and imaginary parts of the
$I=1/2$ mass function $m^2(s)$, together with that for $I=1$, the parameters
of which are also implicitly constrained by Eq.~(\ref{K}). This fixes the
bare $n{\overline n}$ state to be $\sim 1420$ MeV.
It is only  the pole in a propagator that fixes the  parameters
of the state and, of course, it is only
this pole position that is process independent.
  Consequently, we have to continue Eqs.~(\ref{Re},\ref{Im},\ref{P}) 
into the complex $s$--plane onto the appropriate unphysical sheets.  
The scalars being  generally broad and the effect of thresholds
marked, the mass functions vary strongly as one moves
into the complex $s$--plane. 
The sheets are specified by the signs of the ${\rm Im} k_i$, as there is a 
bifurcation at each threshold $i$.
In Table I we give the mass and width $m_{pole}$, $\Gamma_{pole}$ on the
nearest unphysical sheet.

\begin{table}[b]
\vspace{-0.3cm}
\begin{center}
\begin{tabular}{| l | c | c | c | r | c |} 
\hline
\rule[-0.3cm]{0cm}{9mm}  Resonance &  $m_{pole}$& $\Gamma _{pole}$ & sheet \\ 
\hline \hline 
\rule[-0.3cm]{0cm}{9mm} $K_0^*(1430)$ & 1445 & 334 & $-,-,+$  \\ 
\hline 
\rule[-0.3cm]{0cm}{9mm} $\;a_0(980)$  & 1082 & 309 & $-,+,+$  \\ 
\hline 
\rule[-0.3cm]{0cm}{9mm} $\;f_0(980)$  & 1006 &  54 & $-,+,+$  \\ 
\hline 
\rule[-0.3cm]{0cm}{9mm} $\;f_0(1300)$ & 1203 & 361 & $\;-,-,-$\\ 
\hline 
\end{tabular}
\caption{Masses and widths of the scalar nonet members in MeV, 
 with $\gamma = 1.15$.  The sheets are defined by the 
signs of ${\rm Im}k_i$ for $i=1,2,3$, as appropriate.} 
\end{center}
\end{table}
\begin{figure}[t]
\vspace{-0.8cm}
\begin{center}
\mbox{~\epsfig{file=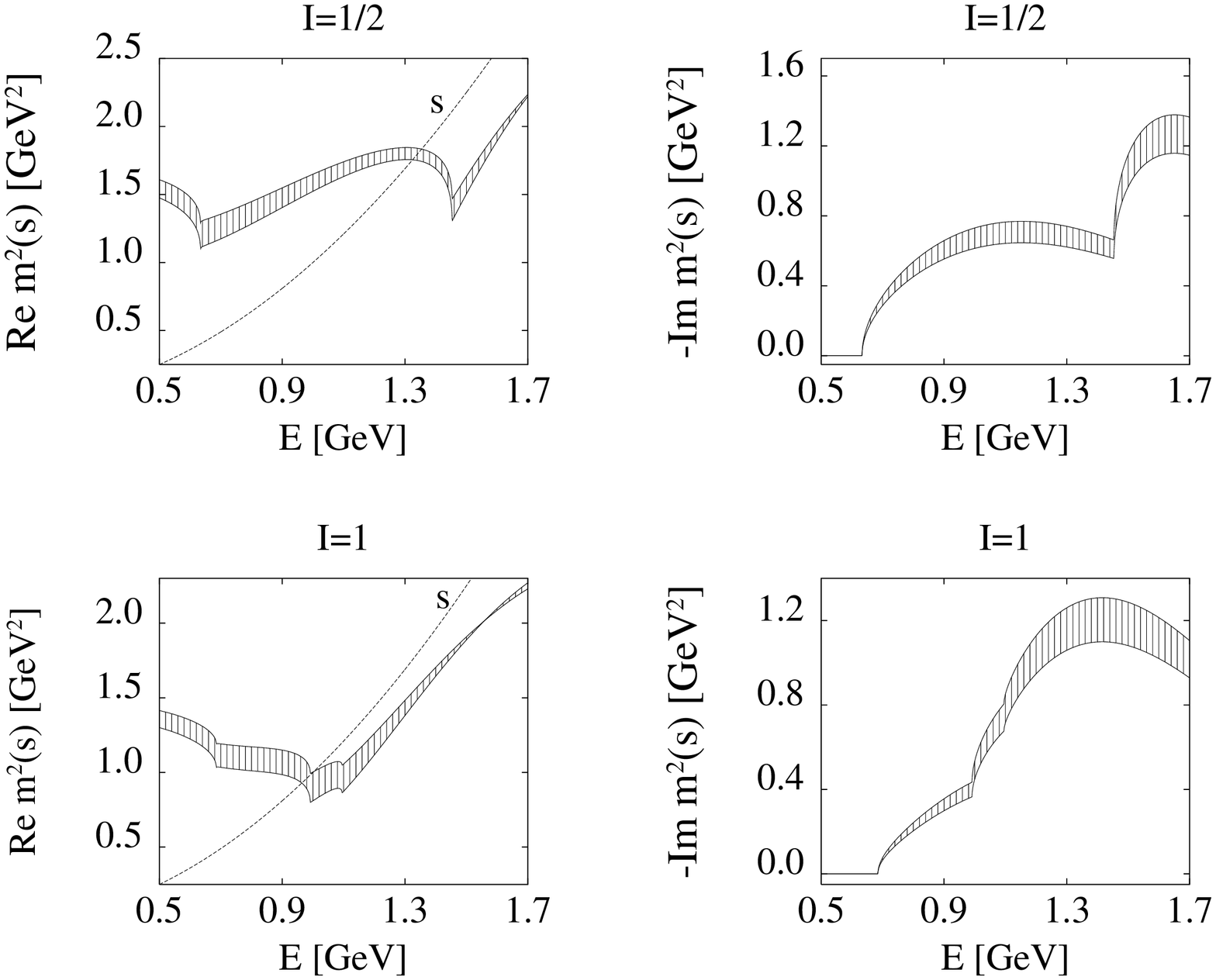,angle=-90,width=13cm}}
\caption{ 
The real and imaginary parts of $m^2(s)$ as functions of 
$E = \protect \sqrt{s}$ for the 
$I=1/2$ and $1$ propagators. Here $m_{0}=1420$ MeV and $\gamma$ is taken to be
between 1.1 and 1.2.  The upper end of each band has $\gamma = 1.1$ 
and the lower $\gamma = 1.2$.}
\end{center}
\end{figure}
%

\vspace{0.3cm}

\noindent {\bf The I=0 sector.} Turning to the $I=0$ sector, the masses
and couplings of the bare
$q{\overline q}$  nonet members are now fixed.  The bare glueball mass
$m_0(gg)$ is taken to be either 1600 MeV from Morningstar and Peardon 
\cite{morn} or 1740 MeV from the {\it IBM} group \cite{ibm}. 
Unquenching by first turning on the pseudoscalar loops of Fig.~1 gives the
complex mass function shown in Fig.~3.  $I_0$, $II_0$ correspond to the pure 
glueball having, respectively, either $SU(3)_f$ singlet couplings or the 
pattern computed
by the {\it IBM} group that favors the higher mass channels. It is by
making 
the imaginary part
of this mass function correspond to a width of $\sim100$ MeV at the
gluestate 
mass
that fixes the coupling $\gamma'$.  There are, of course, analogous mass 
functions for the 
$n{\overline n}$
and $s{\overline s}$ quark model scalars. To find the physical hadrons, 
we have to diagonalize the $ 3 \times 3$ mass matrix
formed from the $n{\overline n}$, $s{\overline s}$ and $gg$ states. 
In physical terms, this allows the quark and glue configurations
to mix through their common communicating channels~:
$\pi\pi$, $K{\overline K}$, $\eta\eta$, $\eta\eta'$ and $\eta'\eta'$. 
Importantly, as noted earlier by Tornqvist \cite{torn}, with the parameters
of the quark multiplet fixed largely by the $K^*_0(1430)$, the
ground state  isoscalars  are naturally the $f_0(980)$ and $f_0(1300)$.
A ground state $s{\overline s}$ scalar up at 1700 MeV~\cite{AmsClose} is
alien to the non-perturbative mixing computed here. The resulting $I=0$
pole positions are given in Table II, again
for central values of the parameters. The presence of the
gluestate has little effect on these predominantly $q{\overline q}$ states.  

\noindent However, as seen in Fig.~3, the mixing has an appreciable effect
on 
the gluestate.  Its coupling
to the $n{\overline n}$ and $s{\overline s}$ states dramatically reduces the
width of the unquenched hadron --- by how much depends on its mass and
coupling pattern.  For an
underlying flavor singlet, the width is down from 100 MeV to $\sim30$ MeV if 
$m_0(gg)=1740$ MeV and to only a few MeV if $m_0(gg)=1600$ MeV.  

\noindent With this suppression of the couplings to the decay channels, the real
part of the mass function, labelled $I$, $II$ in Fig.~3 becomes almost 
independent of $s$. 
The suppression is most appreciable for a lighter
glueball with $SU(3)_f$ couplings.  This is because the mixing with
quark states occurs most through $\pi\pi$ and $K{\overline K}$ intermediate
states.
The {\it IBM} pattern of couplings favors the heavier pseudoscalars and the
width suppression is consequently less.  With the bare 
$n{\overline n}$ state
being at 1420 MeV and the $s{\overline s}$ at 1620 MeV, a gluestate at 1600 MeV
interacts quite differently from one with a bare mass of 1740 MeV.
In Table III we give the corresponding widths of the mainly gluish hadron to
each two pseudoscalar channel.  Notice that there is no large width to
$K{\overline K}$.  Indeed, it is the $\eta\eta$ decay mode that provides the
largest width. These are the predictions that experiment has to check.

\noindent Experiment in fact delivers two potential candidates for the 
unquenched glueball : 
$f_0(1500)$ with a width of $(120 \pm 19)$ MeV seen by the Crystal Barrel 
experiment 
in $p{\overline p}$ annihilation in several different channels \cite{cbar},
and the
$f_J(1710)$ with even spin and width of $(175 \pm 9)$ MeV, first
identified in $J/\psi$ radiative decays \cite{fj}. Our results show that
there is {\it prima facie} difficulty in identifying either of these with
the glueball.
Notice that the  coupling pattern computed by 
the {\it IBM} group~\cite{ibm} for a 
quenched (non-decaying) glueball does not survive strong 
mixing with the other scalars and their decay channels.
\begin{figure}[hp]
\begin{center}
\mbox{~\epsfig{file=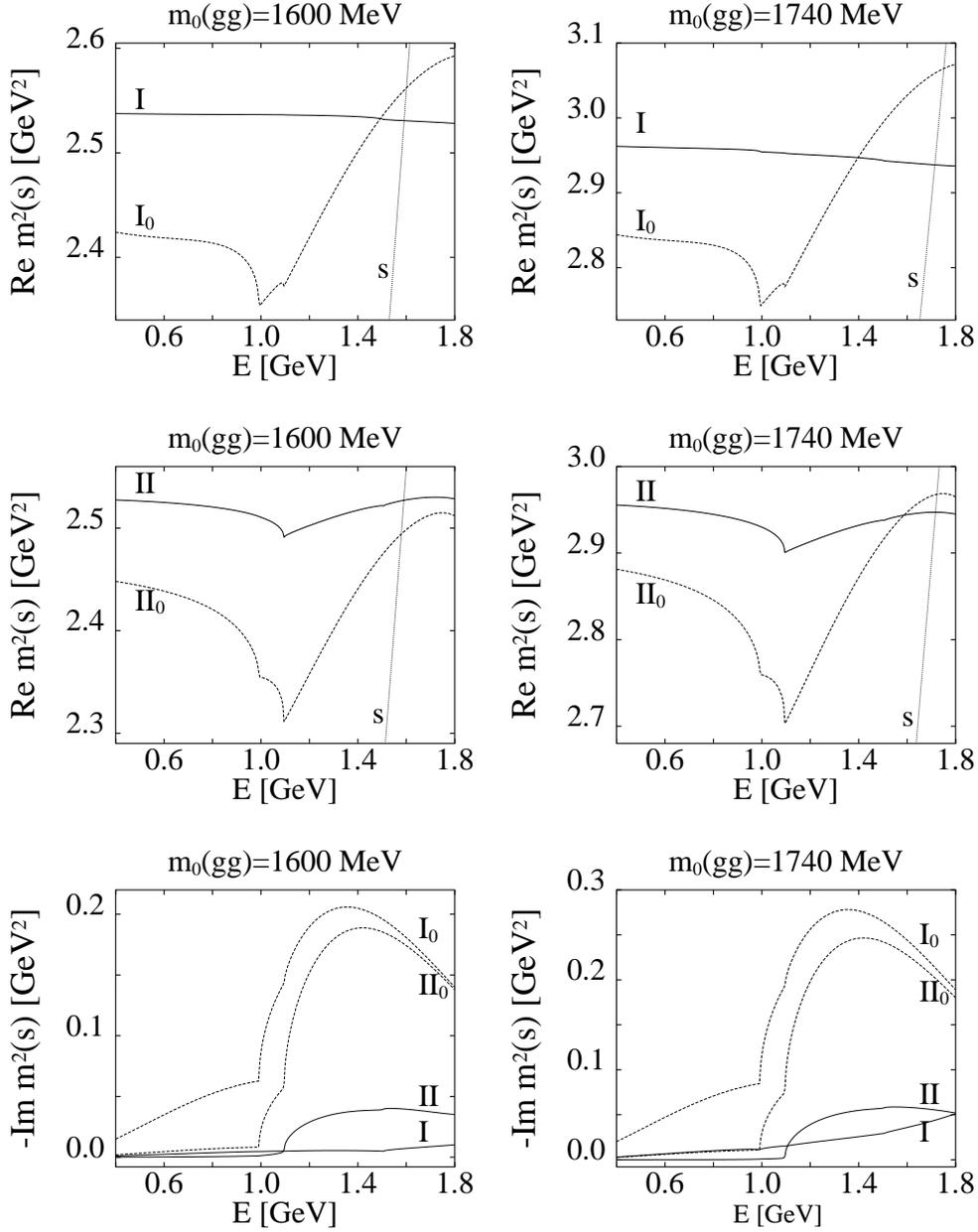,angle=0,width=13cm}}
\caption{
The real and imaginary parts of the glue-state  $m^2(s)$
as functions of $E = \protect \sqrt{s}$
for  two values of bare-mass $m_0(gg) = 1600, 1740$ MeV, suggested by 
lattice calculations. The curves $(I_0)$ and $(I)$ are obtained by using 
$SU(3)_f$ singlet couplings and correspond to the undiagonalized and  
diagonalized cases respectively. The curves $(II_0)$ and $(II)$ are the
analogs obtained by using the pattern of couplings found by the {\it IBM} 
group. 
Here there is negligible  difference  between the curves
with  $\gamma = 1.1$ and $1.2$.}
\end{center}
\vspace{-0.6cm}
\end{figure}
The resulting hadron, rather than coinciding with the $f_0(1710)$ of
Lindenbaum and Longacre~\cite{lind},
has a large $\eta\eta$ decay width  much more like the GAMS $G(1590)$ state 
\cite{gams}.
The latest PDG tables~\cite{PDG} identify this with the WA91 
$f_0(1450)$~\cite{Kirk} and 
the Crystal Barrel $f_0(1500)$~\cite{landua}, 
despite largely contradictory decay information --- agreed with relatively
small statistics from GAMS. 

\begin{table}[t]
\vspace{-0.1cm}
\begin{center}
\begin{tabular}{| l | c | c | c | c | c | c | }
\hline \rule[-0.3cm]{0cm}{9mm}
Couplings& $m_{0}(gg)$ &  $m_{pole}$ & $\Gamma _{pole}$ & sheet \\  
\hline \hline \rule[-0.3cm]{0cm}{9mm}
$SU(3)_f$ singlet & 1600 & 1591 &  3 &  $-,-,+$ \\  
\hline \rule[-0.3cm]{0cm}{9mm}
$\;SU(3)_f$ singlet & 1740 & 1715 & 32 &  $-,-,+$ \\ 
\hline \rule[-0.3cm]{0cm}{9mm}
$\;IBM $            & 1600 & 1589 & 22 &  $-,-,-$ \\
\hline \rule[-0.3cm]{0cm}{9mm}
$\; IBM $           & 1740 & 1715 & 28 &  $-,-,-$ \\
\hline \rule[-0.3cm]{0cm}{9mm}
$\;IBM$ $+$ $4\pi$ channel & 1600 & 1564 & 108 &  $-,+,-,-$ \\
\hline \rule[-0.3cm]{0cm}{9mm}
$\;IBM$ $+$ $4\pi$ channel & 1740 & 1706 & 127 &  $-,-,-,-$ \\
\hline
\end{tabular}
\caption{Mass and width of the glue-state in MeV corresponding to 
various choices of bare mass and couplings. The sheets are defined 
as in Table I.}
\end{center}
\vspace{-0.1cm}
\end{table}

\noindent There is little doubt that the opening of 
multi-pion channels in the 1400 MeV region, which we have so far neglected, 
can  have a marked effect on the gluestate, while changing the ground state 
quark states rather little. To mimic these decays, we can arrange for a  
$4\pi$ channel~\cite{landua} to enhance the dressed gluestate to a 
$\sim 120$ MeV total width. 
This leaves its partial widths to two pseudoscalars as in Table~III. Then the 
larger couplings shift the mass of 
the \lq\lq unquenched" glueball downwards and this 
becomes much more like the $f_0(1500)$ with poles as in Table~II. 
\begin{table}[ht]
\vspace{-0.1cm}
\begin{center}
\begin{tabular}{| l | c | c | c | c | c | c |} 
\hline
\rule[-0.3cm]{0cm}{9mm} Couplings& $m_{0}(gg)$ & $\Gamma _{tot}$ &  
$\Gamma _{\pi \pi}$ & $\Gamma _{K \overline K}$ &  $\Gamma _{\eta \eta}$ &
$\Gamma _{\eta \eta '}$  \rule{0cm}{-4mm} \\ \hline 
\hline \rule[-0.3cm]{0cm}{9mm}
$SU(3)_f$ singlet & 1600 & 4  & 2   & 0.2 & 0.5 & 1.5   \\
\hline \rule[-0.3cm]{0cm}{9mm}
$\;SU(3)_f$ singlet & 1740 & 26 & 5   & 6   & 9   & 6   \\
\hline \rule[-0.3cm]{0cm}{9mm}
$\;IBM$             & 1600 & 25 & 0.5 & 1.5 & 21  & 2   \\
\hline \rule[-0.3cm]{0cm}{9mm}
$\;IBM$             & 1740 & 32 & 0.1 & 1   & 26  & ~5  \\
\hline
\end{tabular}
\caption{Widths of the glue-state to each single two pseudoscalar 
         channel, for various choices of bare mass and couplings.}
\end{center}
\vspace{-0.1cm}
\end{table}

\vspace{0.3cm}

\noindent {\bf Conclusions.} The present work yields definite predictions for 
the decay pattern of the dressed glueball to be compared with experiment. 
The analysis of experiment is however not without its ambiguities. 
Consequently, the challenge 
is to perform a consistent analysis of data on all of peripheral and 
central production, $J/\psi$ radiative decays and $p \overline p$ 
annihilation \cite{mark} and show which of the predictions of Table III is 
in best agreement with experiment. Only then can one claim to have 
discovered the lightest glueball. 

\vspace{0.3cm}

\noindent {\bf Acknowledgements.}
We would like to thank Frank Close and David Morgan for encouraging 
discussions and we acknowledge the support of the 
EURODA$\Phi$NE Network grant ERBCMRXCT920026 of the EC Human 
and Capital Mobility program. 



\end{document}